\documentclass[conference]{IEEEtran}
\IEEEoverridecommandlockouts
\usepackage{amsmath,amssymb,amsfonts}
\usepackage{algorithmic}
\usepackage{graphicx}
\usepackage{textcomp}
\usepackage{xcolor}
\usepackage{array}
\usepackage{cellspace} 
\usepackage{tcolorbox}
\tcbuselibrary{skins}
\tcbuselibrary{breakable}
\usepackage{fancyvrb}
\usepackage{tabularx}
\usepackage{url}
\usepackage[utf8]{inputenc}
\usepackage{array}
\usepackage{setspace}
\usepackage{float}
\usepackage{caption}
\usepackage{subcaption}
\usepackage{wrapfig}
\usepackage{float}

\usepackage[numbers,square,sort&compress]{natbib}

\def\BibTeX{{\rm B\kern-.05em{\sc i\kern-.025em b}\kern-.08em
    T\kern-.1667em\lower.7ex\hbox{E}\kern-.125emX}}

\newtcolorbox[
    use counter=promptcounter
]{prompt}[2][]{
    label={prompt:#1},
    colback=gray!10,
    colframe=gray!50,
    enhanced,
    top=2mm,
    bottom=2mm,
    title=Prompt \thetcbcounter: #2,
    fonttitle=\bfseries,
    before skip=6pt,
    after skip=6pt,
    box align=top,
    float=htb,
    width=\columnwidth
}

\newcounter{mkexamplecounter}

\newtcolorbox[
    use counter=mkexamplecounter
]{mkexample}[2][]{
    label={mkexample:#1},
    colback=gray!10,
    colframe=gray!50,
    enhanced,
    top=1mm,
    bottom=1mm,
    left=1mm,
    right=1mm,
    title={\footnotesize\sffamily Example \themkexamplecounter: #2},
    fonttitle=\bfseries,
    before skip=0pt,
    after skip=0pt,
    box align=top,
    float=tbh,
    fontupper={\scriptsize\sffamily\setstretch{1.2}},
    width=\columnwidth
}

\newtcolorbox{biasdefinition}[1][]{%
    colback=gray!10,
    colframe=gray!50,
    title={\footnotesize\sffamily Box 1: Cognitive Biases and Their Definitions},
    enhanced, 
    top=1mm,
    bottom=1mm,
    left=1mm,
    right=1mm,
    fonttitle=\bfseries,
    fontupper={\scriptsize\sffamily},
    before skip=0pt,
    after skip=0pt,
    box align=top,
    float=tbh,
    #1 
}

\begin{document}

\title{Evaluating Node-tree Interfaces for AI Explainability}

\author{
\IEEEauthorblockN{Lifei Wang}
\IEEEauthorblockA{\normalsize\textit{BTP Innovation} \\ \textit{SAP SE}\\ Palo Alto, USA \\ lifei.wang@sap.com}
\and
\IEEEauthorblockN{Natalie Friedman}
\IEEEauthorblockA{\normalsize\textit{BTP Innovation} \\ \textit{SAP}\\ New York, USA \\ natalie.friedman@sap.com}
\and
\IEEEauthorblockN{Chengchao Zhu}
\IEEEauthorblockA{\normalsize\textit{BTP Innovation} \\ \textit{SAP}\\ Palo Alto, USA \\ chengchao.zhu@sap.com}
\and
\IEEEauthorblockN{Zeshu Zhu}
\IEEEauthorblockA{\normalsize\textit{BTP Innovation} \\ \textit{SAP}\\ Palo Alto, USA \\ zeshu.zhu@sap.com}
\and
\IEEEauthorblockN{S. Joy Mountford}
\IEEEauthorblockA{\normalsize\textit{BTP Innovation} \\ \textit{SAP}\\ Palo Alto, USA \\ joy.mountford@sap.com}
}

\maketitle

\begin{abstract}
As large language models (LLMs) become ubiquitous in workplace tools and decision-making processes, ensuring explainability and fostering user trust are critical. Although advancements in LLM engineering continue, human-centered design is still catching up—particularly when it comes to embedding transparency and trust into AI interfaces. This study evaluates user experiences with two distinct AI interfaces— Node-tree interfaces and chatbot interfaces—to assess their performance in exploratory, follow-up inquiry, decision-making, and problem-solving tasks. Our design-driven approach introduces a Node-tree interface that visually structures AI-generated responses into hierarchically organized, interactive nodes, allowing users to navigate, refine, and follow-up on complex information. In a comparative study with n=20 business users, we observed that while the chatbot interface effectively supports linear, step-by-step queries, it is the Node-tree interface that enhances brainstorming. Quantitative and qualitative findings indicate that Node-tree interfaces not only improve task performance and decision-making support but also promote higher levels of user trust by preserving context. Our findings suggest that adaptive AI interfaces—capable of switching between structured visualizations and conversational formats based on task requirements—can significantly enhance transparency and user confidence in AI-powered systems. This work contributes actionable insights to the fields of human-robot interaction and AI design, particularly for enterprise applications where trust-building is critical for teams. 

\end{abstract}


\section{Introduction}

As Large Language Models (LLMs) become increasingly integrated into workplace tools and decision-making processes, concerns regarding AI  transparency have emerged as a challenge to user trust. The chatbot interface does not lend itself well for source tracking and information organization. Additionally, validating a source or asking for a new information organization can break the flow of conversation. While technical advancements continue to improve LLM accuracy, we see interface design as an opportunity to improve user understandability of AI-generated content.

To address these challenges, this study explores a design-driven approach to improving AI explainability through a Node-tree interface that visually structures AI-generated responses. By organizing information hierarchically in interconnected nodes, we hope that a Node-tree interface could improve information navigation, offering an alternative to traditional chatbot-based interactions. In addition, we intend to understand the impact of novel interfaces (i.e., Node-tree) on trust and performance. We believe this work is valuable for future AI interface development.

To evaluate the Node-tree interface, we conducted a comparative user study focused on three key factors: 
\begin{enumerate}
    \item \textbf{User Trust} – Does the mind map interface enhance confidence in AI responses?
    \item \textbf{Task Performance} – Does it improve user efficiency, accuracy, and decision-making?
    \item \textbf{Interface Usability } – Does the structured visualization provide a more intuitive experience than a traditional chatbot?
\end{enumerate}

By examining how Node-tree interfaces influence user interactions with AI-generated responses, this research provides actionable insights into designing more explainable, trustworthy, and user-friendly AI systems. Our findings not only contribute to the field of human-robot interaction (HRI), but also, we are investigating tasks in enterprise applications (i.e., AI-assisted knowledge management and decision-support systems).

\section{Literature Review}

Explainability bridges the ``communication gap between complex machine functionalities and humans'' \cite{explainability}. For intuitive interaction with an autonomous system, designers must try to build a clear mental model of a machine's intentions, actions, and decision-making rationale.

An explainable interaction design pattern for an AI agent is described by \citet{de2017people}. They describe that if an AI agent explains itself in a way that people find comprehensible, people are more likely to form correct mental models and trust in the system. \citet{bensch2023mining} indicated that AI systems should adapt their behavior to become understandable for users. \citet{edmonds2019tale} also found that comprehensive and real-time visualizations of an AI system's internal decisions were more effective in promoting human trust than explanations just based on summary text descriptions. We suggest that especially in scenarios in which AI agents are facilitating the decision-making process, a Node-tree visualization is a more adaptive model to facilitate the human thinking process. 

\citet{han2021need} indicated that people prefer AI behavior to be more fully explained, that agents need to provide concise summaries, and that agents need to be able to respond to follow-up questions. Based on this work, in our design of interactive features for the Node-tree, our users can ask follow-up questions to any node in the node tree, to enhance their learning about a certain response.

Some current AI tools already use a node tree. These are summarizing tools that help break long-form information down. However, while they provide great features to generate and organize information, they do not allow users to go deeper and take further actions upon the nodes. Such additional actions could be asking follow-up questions about specific nodes; expanding, eliminating, and regenerating a branch or node; or providing a multi-factor analysis to help their overall decision making. Tools that currently exist include: Mapify \cite{mapify}, built for multi-media format information breakdown, Xmind \cite{xmind}, made for everyday brainstorming, and Think Machine \cite{thinkmachine}, which creates a 3D connected net of information. Mymap.ai \cite{mymap} supports most of the actions mentioned above, but this tool is not specialized for the business sector. It is not set up for high risk, profit and cost related business decision-making tasks. The impact financially and socially for the business sector could be significant and valuable.

\section{Methods}

We conducted a between-subjects study to compare user experiences with a structured Node-tree interface and a traditional chatbot. Participants from North America and Europe each completed a variety of tasks in the business sector. We collected quantitative and qualitative data on user trust, task performance, and interface usability. This section outlines our study design, participants, interface features, tasks, and our data collection.

\subsection{Study Design}

The study employed a between-subjects design to study the perception of two interfaces: First, the Node-tree interface, which was the experimental group and second, the chatbot interface, which was the control group. We report fewer questions than we asked participants due to the workshop format limitations and our focus only on the topic of explainability. We are reporting on 15 questions.

\subsubsection{Participants}
A total of n=20 participants (6 male, 14 female) were recruited from Respondent.io \cite{respondent}, with 10 from North America and 10 from Europe. Ages ranged from 23-59 years, m=45.3. These participants were divided into two groups of 10, each assigned to evaluate a different interface design. Participants came from diverse industries including business, consulting, finance, real-estate, information technology, creative services, marketing, hospitality, and fashion. Participants were general business users who could potentially benefit from AI-driven decision-support tools. All participants had prior experience of using GenAI tools (e.g., ChatGPT, Claude) but were not AI experts.

\renewcommand{\thesubfigure}{\roman{subfigure}}
\begin{figure*}[ht!]
    \centering
    \setlength{\tabcolsep}{5pt} 
    \renewcommand{\arraystretch}{1.0} 
    
    \begin{tabular}{ m{0.48\textwidth} m{0.48\textwidth} }
        \centering
        \includegraphics[width=\linewidth]{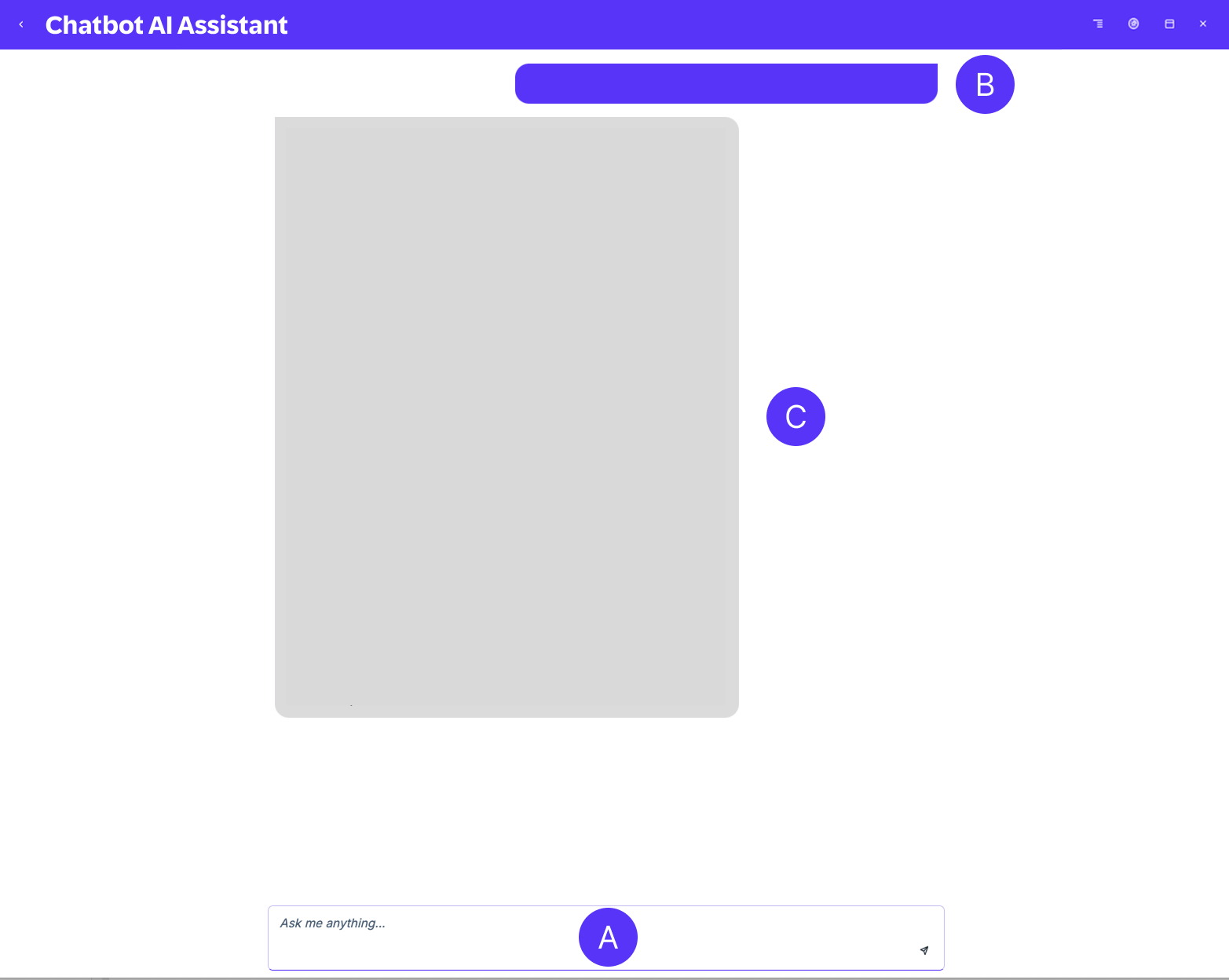}  & 
        \includegraphics[width=\linewidth]{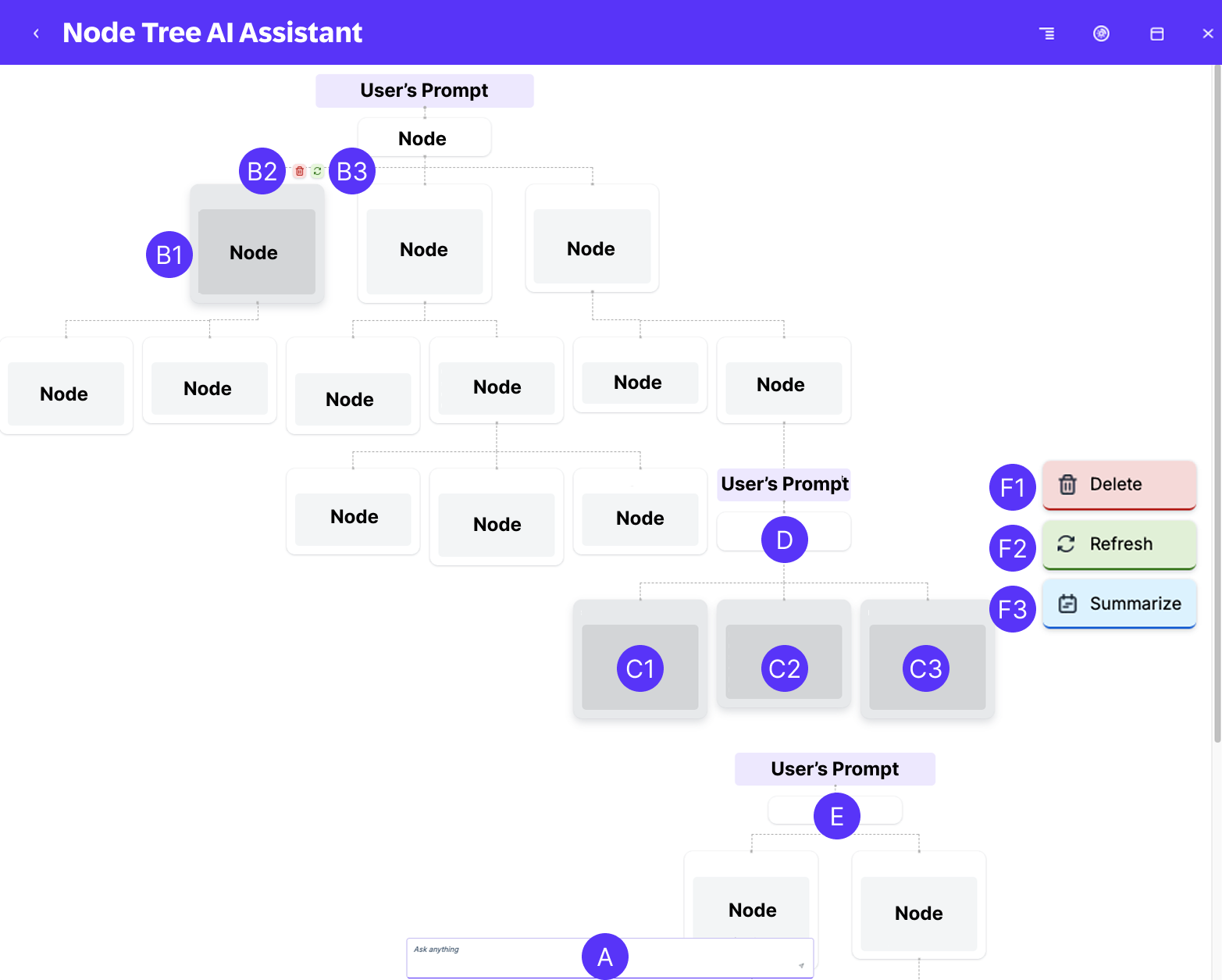}  \\
        
        \centering \textit{(i)} Chatbot interface including the following features: \\ (A) User prompt input box, (B) User prompt box, (C) Chatbot reply output. &
        \centering \textit{(ii)} Node-tree interface including the following features: (A) User prompt box, (B) Node operations – B1: Selected single node, B2: Deletable single node, B3: Refreshable single node, (C) Multi-selected nodes – C1, C2, C3, (D) User's follow-up question, (E) User's follow-up question, (F) Multi-node operations – F1: Multi-node delete, F2: Multi-node refresh, F3: Multi-node summarize. 
    \end{tabular}
    
    \caption{Comparison of the Chatbot interface and the Node-tree interface, showcasing their respective interactive features.}
    \label{fig:comparison}  
\end{figure*}

\subsubsection{Interface Design}
Both interfaces currently only provide text interaction. The Chatbot interface (see Figure \ref{fig:comparison}, (i) was designed based on standard AI chatbots, like ChatGPT. Features included a prompt box (see Figure \ref{fig:comparison}, (i) A), and justified left and right chat bubbles (see Figure \ref{fig:comparison}, (i) B, C). The Node-tree interface features included the following. Firstly, the hierarchical structure showed AI-generated responses broken down into parent nodes and its child nodes below. Complex responses could have multiple levels of nodes and sub-nodes (for visualization, see Figure \ref{fig:comparison} (ii) B1, C1-C3, D, E). Secondly, summary generation gave users the opportunity to select up to five nodes to generate a summary. For visualization of a selection of nodes, see Figure \ref{fig:comparison}, (ii) C1-C3; for visualization of the summary button see Figure \ref{fig:comparison}, (ii) F3. Third, the Node-tree interface allowed for follow-up questions, in which users could select any node to type a follow-up question. For visualization, see Figure \ref{fig:comparison}, (ii) C1-C3, D. Lastly, users could regenerate and delete nodes. When users did this, the child nodes would be removed when a parent node was deleted. For visualization of operations above a selected node, see Figure \ref{fig:comparison}, (ii) B2-B3; for visualization of operations on multi-nodes, see Figure \ref{fig:comparison} (ii), F1-F2.

\subsubsection{Tasks}
In both interface conditions, participants completed four tasks, each simulating real-world scenarios to evaluate how the Node-tree interface influenced user's performance:
\begin{enumerate}
    \item \textbf{Exploratory Task}: Participants asked the AI assistant for help with planning an international trip.

    \item \textbf{Follow-Up Inquiry Task:} Participants were asked to follow up on their previous text conversation with an AI assistant to identify cost-saving strategies for international flights. 

    \item \textbf{Decision-Making Task:} Participants compared three potential locations for a new city park and then made a choice relying on AI-generated text.

    \item \textbf{Problem-Solving Task:} Participants determined the optimal time and location for a food distribution event. More specifically, they were asked, ``You're helping a nonprofit in your city organize a food distribution event. You need to determine the best time and location for maximum attendance while minimizing costs. Ask the AI assistant to help you determine the best time and location for a food distribution event."

\end{enumerate}

\subsubsection{Measures}
We assessed participants on both quantitative and qualitative metrics after completing each task and at the end of the study.

After each task, participants provided ratings on a 10-point Likert scale or multiple choice answers:
\begin{itemize}
    \item \textbf{User Trust:} ``How much did you trust the AI Assistant's responses?" (1 = Not confident, 10 = Very confident)
    \item \textbf{Task Performance:} ``Did the Node-tree help you understand the problem better?" (Options: Highly helpful/ Somewhat helpful/ Neither helpful nor unhelpful/ Somewhat unhelpful/ Highly unhelpful)
    \item \textbf{Interface Usability:} ``Was the Node-tree easy to navigate?" (Options: Very easy / Somewhat easy / Neither easy nor difficult / Somewhat difficult / Very difficult)
 
\end{itemize}

To analyze this data, we conducted t-tests for both conditions. For the overall feedback section, we gathered final reflections by asking, ``Tell us about your experience using the AI assistant. What worked well? What could be improved?”

This combination of quantitative and qualitative data was used to determine whether a structured Node-tree interface enhanced usability compared to a traditional chatbot interface.

\section{Results}

Across all tasks, both interfaces were evaluated on user trust, task performance, and interface usability. At the beginning of the study, we also asked about challenges with current AI tools. See below for our results. 

\subsection{Challenges with AI Tools} 

Before participants interacted with our tool, participants across both groups cited recurring challenges with existing AI tools. Firstly, they mentioned accuracy and reliability issues; Users expressed concerns about incorrect or misleading responses. Next, they also complained about lack of organizational features in current conversational interfaces. Lastly, people mentioned that they wanted to ask follow-up questions, but they were hesitant to ask these questions to avoid derailing the conversation. While we only addressed some of these issues via our Node-tree interface, we hope that these results can be used by designers to improve the usability of AI agents too. 

\subsection{User Trust}
Trust is difficult to measure and is both user and use-case dependent. However, we found that 90\% of the Node-tree users reported that the Node-tree influenced their trust in AI responses positively. No participants found that the Node-tree interface reduced level of trust. Furthermore, 80\% of Node-tree participants were positive about using AI-generated suggestions in real-life situations for completing problem-solving tasks. For the same task, 70\% of chatbot users reported positive feedback. While this difference is not significant (t(19)= .21, p=.6), we think that difference in the small sample size is promising for future studies.

\subsection{Task Performance}

In the exploratory and decision-making tasks, the Node-tree group also consistently outperformed the chatbot group. This suggests that users preferred visually interconnected information when evaluating multiple factors or brainstorming. In the problem-solving tasks, the chatbot group demonstrated better performance in structured, step-by-step tasks, indicating that users found linear, conversational interactions more effective in those tasks. We believe this demonstrates that the chatbot interface helped participants do linear thinking, whereas the interface of the Node-tree helped participants think more out of the box. 

More specifically, Node-tree users appreciated the ability to maintain context within the Node-tree interface. One user noted that the Node-tree interface ``[displays] all aspects of [my] question ... and it does it in a nice and summarized way." Conversely, chatbot participants experienced difficulties with losing context during follow-up questions with the chatbot interface. These users specifically pointed out that ``several times, the context is lost and a generic answer is provided."

\subsection{Interface Usability}
When we asked about interface usability, Node-Tree users found the interface's structured format intuitive to navigate. One user noted that the ``mapping was clear, easy to navigate, and provided many factors that I might not have thought about." 90\% of Node-Tree users considered the interface easy to navigate.

In the qualitative feedback, we found that both interface styles are needed. Firstly, for both interfaces, users suggested visual elements to be incorporated, like images and tables to improve information presentation. Noticing that the AI agent often removed important context in its responses, participants suggested features like “suggesting related questions” to help maintain context and keep the conversation on track.

These findings demonstrate the necessity to align AI interface design with user needs. We believe that structured and conversational interfaces serve distinct purposes. We expect future work to build on these insights to further refine all design components iteratively to help with usability, context retention, and easy-to-use information presentation.

\section{Discussion}

In this section, we reflect on the implications of our preliminary findings, providing design recommendations based on user feedback and observed interaction patterns. Given the small sample size, we were surprised that we saw a difference in how the Node-tree interface supports brainstorming, while the chatbot interface was well-suited for single-step prompts. 


\subsection{Design Recommendations}
In brainstorming and decision-making tasks, the Node-tree interface enhances information organization by visually structuring information, making complex information easier to navigate compared to chatbots. Based on our study findings, we suggest the following  interface opportunities to improve explainability.

Our preliminary results, suggest starting to use node trees for decision-making and brainstorming tasks. The interface supports understanding by providing a structured overview of the AI-generated recommendations. From our open-ended responses, we found that Node-trees also helped with idea generation and research by showing interconnected concepts. Using traditional chatbot interfaces seem best suited for single-step questions, when users are looking for instant and straightforward responses. Linear workflows, such as step-by-step process for troubleshooting, could also be supported by chatbot interfaces. To be adaptive, interfaces should respond to the specific question type and update the interface accordingly. This is in line with what \citet{bensch2023mining} suggested, in that AI systems should ``adapt'' their behavior to better match human understanding.

\subsection{Limitations and Future Work}

We note that there were several limitations in this work, both in the study design and in the analysis. Due to the difference in tasks, we had to ask different questions after each task, so they were not directly comparable. As an exploratory study, we were wiling to accept this open-ended approach. In future work, we would use a larger sample size for the identical questions across all conditions. We do not report statistical significance because the sample size was not high enough, and instead we report user data trends.

\section{Conclusion}
This study explored the effectiveness of Node-tree interfaces compared to traditional chatbot interfaces for enhancing AI explainability and user trust. Our findings revealed that interface effectiveness is highly context-dependent: Node-tree interfaces excelled in our complex decision-making and brainstorming tasks. They also demonstrated increased user trust and improved problem-solving capabilities. Chatbot interfaces prove more effective for linear, step-by-step tasks requiring immediate responses. These results suggest that future AI interfaces should be adaptable to users, and try to match presentation format to specific task requirements. While our study was limited by sample size, we surprisingly found a difference in group performance. These results are promising valuable for future interface research and development. Our results are particularly useful in enterprise settings where structured information presentation and trust-building are crucial. We hope to address future work on developing adaptive interfaces that can seamlessly switch between formats based on task type.

\section*{Acknowledgment}

The authors thank SAP for supporting this work and the participants for giving their time to research.

\bibliographystyle{IEEEtranN}   


\end{document}